\documentclass[format=sigconf,screen,authorversion]{acmart}

\usepackage{booktabs} 

\usepackage{cleveref}

\crefformat{footnote}{#2\footnotemark[#1]#3}

\usepackage{balance}

\usepackage{algorithm}
\usepackage[]{algpseudocode}
\makeatletter
\let\OldStatex\Statex
\renewcommand{\Statex}[1][3]{%
  \setlength\@tempdima{\algorithmicindent}%
  \OldStatex\hskip\dimexpr#1\@tempdima\relax}
\makeatother

\algnewcommand{\LeftComment}[1]{\OldStatex \(\triangleright\) #1}

\usepackage{xcolor}
\usepackage{listings}
\usepackage{tabularx}
\usepackage{array,tabu}
\usepackage{siunitx}
\usepackage{multirow}
\usepackage[para]{threeparttablex}

\graphicspath{{figures/}}

\newcommand{\intelphi}{Intel Xeon Phi}

\newcommand{\iphi}{Xeon Phi}

\newcommand{\intelreg}{Intel\textregistered}

\newcommand{\intelphireg}{Intel{\textregistered} Xeon Phi\textsuperscript{TM}}

\newcommand{\iphireg}{Xeon Phi\textsuperscript{TM}}

\newcommand{\abinitio}{\textit{ab initio}}
\newcommand{\second}{2\textsuperscript{nd}}
\newcommand{\bigon}[1]{$\mathcal{O}(N^{#1})$}

\newcolumntype{Y}{>{\centering\arraybackslash}X}

\tabucolumn{Y}
\newcolumntype{Z}{S[table-number-alignment = center,
					table-figures-integer = 3,
                    table-figures-decimal = 0]}
\tabucolumn{Z}

\copyrightyear{2017} 
\acmYear{2017} 
\setcopyright{usgovmixed}
\acmConference{SC17}{November 12--17, 2017}{Denver, CO, USA}
\acmPrice{15.00}
\acmDOI{10.1145/3126908.3126956}
\acmISBN{978-1-4503-5114-0/17/11}

\begin{document}
\title[MPI/OpenMP parallelization of the Hartree-Fock method]{An efficient MPI/OpenMP parallelization of the Hartree-Fock method for the second generation of \intelphireg\ processor}

\author{Vladimir Mironov}
\affiliation{
	\institution{Lomonosov Moscow State University}
    \streetaddress{Leninskie Gory 1/3}
    \city{Moscow}
    \postcode{119991}
    \country{Russian Federation}
} \email{vmironov@lcc.chem.msu.ru}

\author{Yuri Alexeev}
\affiliation{
	\institution{Argonne National Laboratory, Leadership Computing Facility}
    \city{Argonne}
    \state{Illinois}
    \postcode{60439}
    \country{USA}
} \email{yuri@alcf.anl.gov}

\author{Kristopher Keipert}
\affiliation{
	\institution{Department of Chemistry and Ames Laboratory, Iowa State University}
    \city{Ames}
    \state{Iowa}
    \postcode{50011-3111}
    \country{USA}
} \email{kris@si.msg.chem.iastate.edu}

\author{Michael D'mello}
\affiliation{
	\institution{Intel Corporation}
    \city{Schaumburg}
    \state{Illinois}
    \postcode{60173}
    \country{USA}
} \email{michael.dmello@intel.com} 

\author{Alexander Moskovsky}
\affiliation{
	\institution{RSC Technologies}
    \city{Moscow}
    \country{Russian Federation}
} \email{moskov@rsc-tech.ru}    

\author{Mark S. Gordon}
\affiliation{
	\institution{Department of Chemistry and Ames Laboratory, Iowa State University}
    \city{Ames}
    \state{Iowa}
    \postcode{50011-3111}
    \country{USA}
} \email{mgordon@iastate.edu}

\renewcommand{\shortauthors}{V. Mironov et al.}

\begin{abstract}
Modern OpenMP threading techniques are used to convert the MPI-only Hartree-Fock code in the GAMESS program to a hybrid MPI/OpenMP algorithm. Two separate implementations that differ by the sharing or replication of key data structures among threads are considered, density and Fock matrices. All implementations are benchmarked on a super-computer of 3,000 \intelphireg\ processors. With 64 cores per processor, scaling numbers are reported on up to 192,000 cores. The hybrid MPI/OpenMP implementation reduces the memory footprint by approximately 200 times compared to the legacy code. The MPI/OpenMP code was shown to run up to six times faster than the original for a range of molecular system sizes.
\end{abstract}

%
%



\maketitle

\section{Introduction}
The field of computational chemistry encompasses a wide range of empirical, semi-empirical, and \abinitio\ methods that are used to compute the structure and properties of molecular systems. These methods therefore have a significant impact on not only chemistry, but materials, physics, engineering and the biological sciences as well. \textit{Ab initio} methods are rigorously derived from quantum mechanics. In principle, \abinitio\ methods are more accurate than methods with empirically fitted parameters. Unfortunately, this accuracy comes at significant computational expense. For example, the time to solution for Hartree-Fock (HF) and Density Functional Theory (DFT) methods scale as approximately \bigon{3}, where N is the number of degrees of freedom in the molecular system. The HF solution is commonly used as a starting point for more accurate \abinitio\ methods, such as second order perturbation theory and coupled-cluster theory with single, double, and perturbative triple excitations. These post-HF methods scale as \bigon{5} and \bigon{7}, respectively. These computational demands clearly require efficient utilization of parallel computers to treat increasingly large molecular systems with high accuracy.

Modern high performance computing hardware architecture has substantially changed over the last 10 to 15 years. Nowadays, a ``many-core'' philosophy is common to most platforms. For example, the Intel Xeon Phi processor can have up to 72 cores. For good resource utilization, this necessitates (hybrid) MPI+X parallelism in application software.

The subject of this work is the successful adaptation of the HF method in the General Atomic and Molecular Electronic Structure System (GAMESS) quantum chemistry package to the second-generation Intel Xeon Phi processor platform. GAMESS is a free quantum chemistry software package maintained by the Gordon research group at Iowa State University \cite{gordon2005advances}. GAMESS has been cited more than 10,000 times in the literature, downloaded more than 30,000 times and includes a wide array of quantum chemistry methods. The objective here is to start with the MPI-only version of GAMESS HF and systematically introduce optimizations which improve performance and reduce the memory footprint. Many existing methods in GAMESS are parallelized with MPI. OpenMP is an attractive high-level threading application program interface (API) that is scalable and portable. The OpenMP interface conveniently enables sharing of the two major objects in the HF self-consistent field (SCF) loop: the density matrix and the Fock matrix.

The density and Fock data structures account for the majority of the memory footprint of each MPI process. Indeed, since these two objects are replicated across the MPI processes, memory capacity limits can easily come into play if one tries to improve the time to solution using a large number of cores. By sharing one or both of the aforementioned objects between threads, one can reduce the memory footprint and more easily leverage all of the resources (cores, fast memory etc.) of the Intel Xeon Phi processor. Reducing the memory footprint is also expected to lead to better cache utilization, and, therefore, enhanced performance. Two hybrid OpenMP/MPI implementations of the publicly available version of the GAMESS (MPI-only) code base were constructed for this work. The first version is referred to as the ``shared density private Fock'', or ``private Fock'' version of the code. The second version is referred to as the ``shared density shared Fock'', or ``shared Fock'' version.

In the following section, a brief survey of related work is presented. Next, key algorithmic features of the HF self-consistent field (SCF) method are discussed. Then, a description of the computer hardware test bed that was used for benchmarking purposes is presented. An explanation of the code transformations employed in the hybrid implementation in this work follows. Next, the memory and time-to-solution results of the hybrid approach are shown. Results on up to 3,000 Intel Xeon Phi processors are presented for a range of chemical system sizes. The work ends with concluding remarks and a discussion of directions for future work.

\section{Related work}
\label{sec:related}
The HF algorithm has been a primary parallelization target since the onset of parallel computing. The primary computational components of the HF algorithm are construction of the density and Fock matrices, that are described in \cref{sec:hf} of this work. The irregular task and data access patterns during Fock matrix construction bring significant challenges to efficient parallel distribution of the computation. The poor scaling of Fock matrix diagonalization is a major expense as well. Linear scaling methods like the fragment molecular orbital method (FMO) have been successfully applied to thousands of atoms and CPU cores \cite{alexeev2012heuristic,umeda2010parallel}, but such methods introduce additional approximations \cite{fedorov2007extending,fedorov2009}. In any case, fragmentations methods may benefit from optimizations of the core HF algorithm as well.

Early HF parallelization efforts focused on the distributed computation of the many electron repulsion integrals (ERIs) required for Fock matrix construction via MPI or other message passing libraries. The Fock and density matrices were often replicated for each rank, and load balancing algorithms were a primary optimization target. Blocking and clustering techniques were explored in depth in a landmark paper by \citet{foster1996toward}. Their contributions were implemented in the quantum chemistry package NWChem~\cite{valiev2010nwchem}. In a follow-up paper by \citet{harrison1996toward}, a node-distributed HF implementation was introduced. In this work, both the density and Fock matrices were distributed across nodes using globally addressable array (GA). In a more recent work UPC++ library was used to achieve this goal \cite{ozog2016hartree}. A similar approach was used to implement distributed data parallel HF by \cite{alexeev2002distributed,alexeev2007parallel} in the GAMESS code. This implementation utilizes the Distributed Data Interface (DDI) message passing library \cite{fletcher2000distributed}. To further address the load balancing issues, a work stealing technique was introduced by \citet{liu2014new}.

A detailed study and analysis of the scalability of Fock matrix construction and density matrix construction \cite{chow2015parallel}, including the effects of load imbalance, was explored in a work by \mbox{\citet{chow2015scaling}}. In this work, density matrix construction was achieved by density purification techniques and the resulting implementation was scaled up to 8,100 Tianhe-2 Intel Xeon Phi first generation co-processors. In fact, a number of attempts have been made to design efficient implementations of HF for accelerators \cite{asadchev2012new,chow2015scaling,ufimtsev2008quantum,ufimtsev2009quantum,wilkinson2011acceleration} and other post-HF methods \cite{apra2014efficient}. A major issue in this context is the management of shared data structures between cores -- in particular, the density and Fock matrices. OpenMP HF implementations with a replicated Fock matrix and shared density matrix have been explored in the work of \citet{ishimura2010mpi} and \citet{mironov2015quantum}. The differences between these works are in the workload distribution among MPI ranks and OpenMP threads. The current work borrows some techniques from these previous works which implement HF for accelerators. The result is a hybrid MPI/OpenMP implementation that is designed to scale well on a large number of Intel Xeon Phi processors, while at the same time managing the memory footprint and maintaining compatibility with the original GAMESS codebase.

\section{Hartree-Fock method}
\label{sec:hf}
The HF method is used to iteratively solve the electronic Schr\"odinger equation for a many-body system. The resulting electronic energy and electronic wave function can be used to compute equilibrium geometries and a variety of molecular properties. The wave function is constructed of a finite set of basis functions suitable for algebraic representation of the integro-differential HF equations.  Central to HF is an effective one-electron Hamiltonian called the Fock operator which describes electron-electron interactions by mean field theory. In computational practice, the Fock operator is defined in matrix form (Fock matrix). The HF working equations are then represented by a nonlinear eigenvalue problem called the Hartree-Fock equations: 
\begin{equation}\label{eqn:scf}
	\mathbf{FC}={\epsilon}\mathbf{SC}
\end{equation}
where $\epsilon$ is a diagonal matrix corresponding to the electronic orbital energies, $\mathbf{F}$ is a Fock matrix, $\mathbf{C}$ is matrix of molecular orbital (MO) coefficients, and $\mathbf{S}$ is the overlap matrix of the atomic orbital (AO) basis set. The HF equations are solved numerically by self-consistent field (SCF) iterations.

The SCF iterations are preceded by computation of an initial guess density matrix and core Hamiltonian. An initial Fock matrix is constructed from terms of the core Hamiltonian and a symmetric orthogonalization matrix. Next, the Fock matrix is diagonalized to provide the MO coefficients $\mathbf{C}$. These MO coefficients are used to compute an initial guess density matrix. The SCF iterations follow, in which a new Fock matrix is constructed as a function of the guess density matrix. Diagonalization of the updated Fock matrix provides a new set of MO coefficients which are used to update the density matrix. This iterative process continues until convergence is reached, which is defined by the root-mean-squared difference of consecutive densities lying below a chosen convergence threshold.

Contrary to what one might expect, the most time-consuming part of the calculation is not the solution of the Hartree-Fock equations, but rather the construction of the Fock matrix~\cite{janssen2008}. The calculation of the Fock matrix elements can be separated into one-electron and two-electron components. The computational complexity of these two parts is \bigon{2} and \bigon{4}, respectively. In most cases of practical interest, the calculation of the two-electron contribution to the Fock matrix occupies the majority of the overall compute time.

\section{Optimization and parallelization of the Hartree-Fock method}
\label{sec:optandpar}

\begin{algorithm}[t]
\caption{MPI parallelization of SCF in stock GAMESS}\label{alg:mpi}
\begin{algorithmic}[1]
\For {$i$ = 1, $NShells$}
	\For {$j$ = 1, $i$}
    \State \textbf{call} ddi\_dlbnext($ij$)		\Comment MPI DLB: check I and J indices
    	\For {$k = 1, i$}
        	\State $k$==$i$ ? $l_{max}\gets k$ : $l_{max}\gets j$
            \For {$l$ = 1, $l_{max}$}
            	\LeftComment Schwartz screening:
                \State $screened\gets$ schwartz($i,j,k,l$)
                \If {\textbf{not} $screened$}
                	\State \textbf{call} eri($i,j,k,l,X_{ijkl}$) \Comment Calculate $(i,j|k,l)$
                    \LeftComment Update process-local 2e-Fock matrix:
            		\State $Fock_{ij,kl,ik,jl,il,jk}$ += 
                    \Statex[7] $X_{ijkl}\cdot D_{kl,ij,jl,ik,jk,il}$
                \EndIf
            \EndFor
        \EndFor
    \EndFor
\EndFor
\LeftComment 2e-Fock matrix reduction over MPI ranks:
\State \textbf{call} ddi\_gsumf($Fock$)
\end{algorithmic}
\end{algorithm}

\subsection{General considerations and design}
\label{ssec:general}

In this section, three implementations of the HF algorithm are presented: the original MPI algorithm \cite{schmidt1993general} and two new hybrid MPI/OpenMP algorithms. As mentioned earlier, the most expensive steps in HF are the computation of ERIs and the contribution of ERIs multiplied by corresponding density elements during construction of the Fock matrix. The symmetry-unique ERIs are labeled in four dimensions over $i$, $j$, $k$, $l$ shell\footnote{By term \textit{shell} we mean a group of basis set functions related to the same atom and sharing same set of internal parameters. Grouping basis functions into shells is a common technique in Gaussian-based quantum chemistry codes like GAMESS.} indices. The symmetry-unique quartet shell indices are traversed during Fock matrix construction. Parallelization over the four indices is complicated by the high order of permutational symmetry for shell indices. In addition, many integrals are very small in magnitude and are screened out using the Cauchy-Schwarz inequality equation (see e.q. \citep[p. 118]{janssen2008}). Each ERI is used to construct six elements of the Fock matrix shown in \crefrange{eqn:fockupd:a}{eqn:fockupd:f} where $(i,j|k,l)$ corresponds to a single ERI:

\begin{subequations}
	\label{eqn:fockupd}
	\begin{align}
		F_{ij} &\gets (i,j|k,l)\cdot D_{kl}; \label{eqn:fockupd:a} \\
		F_{kl} &\gets (i,j|k,l)\cdot D_{ij}; \label{eqn:fockupd:b}\\
		F_{ik} &\gets (i,j|k,l)\cdot D_{jl}; \label{eqn:fockupd:c}\\
		F_{jl} &\gets (i,j|k,l)\cdot D_{ik}; \label{eqn:fockupd:d}\\
		F_{il} &\gets (i,j|k,l)\cdot D_{jk}; \label{eqn:fockupd:e}\\
		F_{jk} &\gets (i,j|k,l)\cdot D_{il}; \label{eqn:fockupd:f}
	\end{align}
\end{subequations}

The irregular storage and access of ERIs during Fock matrix construction is a significant computational challenge. Also, the Fock matrix construction is distributed among ranks, and the final Fock matrix is summed up by a reduction. A detailed explanation of the SCF implementation in GAMESS can be found elsewhere~\cite{schmidt1993general}.

\subsection{MPI-based Hartree-Fock algorithm}
\label{ssec:mpi}
The MPI parallelization in the official release of the GAMESS code is shown in \Cref{alg:mpi}. While this implementation has been remarkably successful, it has the disadvantage of a very high memory footprint. This is because a number of data structures (including the density matrix, the atomic orbital overlap matrix, and the one- and two-electron contributions to the Fock matrix) are replicated across MPI ranks. It is a major issue for processors which have a large number of cores (like the Intel Xeon Phi). For example, running~256~MPI ranks on a single Intel Xeon Phi processor increases the memory footprint for both density and Fock matrices by a factor of 256 times. This implementation is therefore severely restricted when it comes to the size of the chemical systems that can be made to fit in memory. 

In a typical calculation, the number of shells (see $NShells$ in \Cref{alg:mpi}) is less than one thousand. Most often, the number can be on the order of a few hundred shells. Thus, parallelization over a two shell indices (\Cref{alg:mpi}) frequently results in  load imbalances. The HF algorithm in GAMESS was originally designed for small- to medium-sized x86 CPU architecture clusters when load balancing is not such a significant issue. However, switching to computer systems with larger parallelism (large number of compute nodes) requires a change of approach for load balancing. Multiple solutions exist for this problem. Perhaps the simplest one is to use more shell indices to increase the iteration space and improve the load balance or introduce multilevel load balancing schemes.

\subsection{Hybrid OpenMP/MPI Hartree-Fock algorithm}
\label{ssec:omp}
In this section, the hybrid MPI/OpenMP two-electron Fock matrix code implementations of the current work are described. The main goal of this implementation is to reduce the memory footprint of the MPI-based code and to improve the load balancing by utilizing the OpenMP runtime library.

Modern computational cluster nodes can have a large number of cores operating on a single random access memory. In order to efficiently utilize all of the available CPU cores, it is necessary to run many threads of execution. The major disadvantage of an MPI-only HF code is that all of the data structures are replicated across MPI processes (ranks) -- since to spawn a process is the only way to use a CPU core. In practice, it is found that the memory footprint gets prohibitive rather quickly as the chemical system is scaled up. It follows from \Cref{alg:mpi} that only the Fock matrix update incurs a potential race-condition (write dependencies) when leveraging multiple threads. Other large memory objects like the density matrix, the atomic orbital overlap matrix, and others do not exhibit this problem, because they are read-only matrices, and as a result they can be safely shared across all threads for each MPI rank.

\begin{algorithm}[t]
\caption{Hybrid MPI-OpenMP SCF algorithm; Fock matrix is replicated across all threads i.e. Fock matrix is private.}\label{alg:omp}
\begin{algorithmic}[1]

\State !\$omp parallel private($j,k,l,l_{max},X_{ijkl}$) shared($I$) 
\Statex[1] reduction($+:Fock$)
\Loop
\State !\$omp master
    \State \textbf{call} ddi\_dlbnext($i$)		\Comment MPI DLB: get new I index
\State !\$omp end master
\State !\$omp barrier
\State !\$omp do collapse(2) schedule(dynamic,1)
	\For {$j$ = 1, $i$}
    	\For {$k$ = 1, $i$}
        	\State $k$==$i$ ? $l_{max}\gets k$ : $l_{max}\gets j$
            \For {$l$ = 1, $l_{max}$}
            	\LeftComment Schwartz screening:
                \State $screened\gets$ schwartz($i,j,k,l$) 
                \If {\textbf{not} $screened$}
                	
                    \State \textbf{call} eri($i,j,k,l,X_{ijkl}$) \Comment Calculate $(i,j|k,l)$
                    \LeftComment Update private 2e-Fock matrix:
            		\State $Fock_{ij,kl,ik,jl,il,jk}$ += 
                    \Statex[7] $X_{ijkl}\cdot D_{kl,ij,jl,ik,jk,il}$
                            
                \EndIf
            \EndFor
            \EndFor
    \EndFor
\State !\$omp end do
\EndLoop
\State !\$omp end parallel
\State \textbf{call} ddi\_gsumf($Fock$) \Comment 2e-Fock matrix reduction over MPI
\end{algorithmic}
\end{algorithm}

In a first attempt, a hybrid MPI/OpenMP Hartree-Fock code was developed with the Fock matrix replicated across threads (\mbox{\Cref{alg:omp}}). This is what is referred to as the private Fock (hybrid) version of the code. In the first loop, the master thread of each MPI rank updates the $i$ index. This operation is protected by implicit and explicit barriers. OpenMP parallelization is implemented over combined $j$ and $k$ shell loops. Joining loops provides a much larger pool of tasks and thereby alleviates any load balancing issues that may arise. To lend credence to this idea, static and dynamic schedules of OpenMP were tested for the collapsed loop. No significant difference between the various OpenMP load balancer modes was observed. The $l$ loop is the same as in the original implementation of GAMESS. The last step is the same as in the MPI-based algorithm: reduction of the Fock matrix over MPI processes.

Sharing all of the large matrices except the Fock matrix saves an enormous amount of memory on the multicore systems. The observed memory footprints on the latest Xeon and Xeon Phi CPUs were reduced about 5 times. However, the ultimate goal of this work is to move all of the large data structures to shared memory.

It is not straightforward to remove Fock matrix write dependencies in the OpenMP region. As shown in \crefrange{eqn:fockupd:a}{eqn:fockupd:f}, up to six Fock matrix elements are updated at one time by each thread. The ERI contribution is added to the three shell column-blocks of the Fock matrix simultaneously -- namely the $i$, $j$, and $k$ blocks. Each block corresponds to one shell and to all basis set functions associated with this shell. The main idea of the present approach is to use thread-private storage for each of these blocks. They are used as a buffer accumulating partial Fock matrix contribution and help to avoid write dependency. Partial Fock matrix contributions are flushed to the full matrices when the corresponding shell index changes.

\begin{algorithm}[t]
\caption{Hybrid MPI-OpenMP SCF algorithm; Fock matrix is shared across all threads.}\label{alg:ompsh}
\begin{algorithmic}[1]

\State $mxsize \gets$ ubound($Fock$)$\cdot shellSize$
\State $nthreads \gets$ omp\_get\_max\_threads()
\State allocate($F_I(mxsize,nthreads, F_J(mxsize,nthreads)$)

\State !\$omp parallel shared($F_I, F_J, Fock$) \&
\Statex[1] private($i,j,k,l,ithread$)
\State $ithread \gets$ omp\_get\_thread\_num()
\Loop
	\State !\$omp master
    
    \State \textbf{call} ddi\_dlbnext($ij$) \Comment MPI DLB: get new combined IJ index
	\State !\$omp end master
	\State !\$omp barrier
    \State $i,j\gets ij$ 			\Comment Deduce I and J indices
    \State ${kl}_{max}\gets i,j$ 	\Comment Deduce KL-loop limit
    \State $screened\gets$ schwartz($i,j,i,j$) \Comment I and J prescreening 
    \If {\textbf{not} $screened$}
    
    	\If {$i \ne i_{old}$} \Comment If $i$ was changed flush $F_I$
    		\State $Fock(:,i)$+=$\sum F_I(:,1$:$nthreads)$
            \State !\$omp barrier
   		\EndIf
    	\State !\$omp do schedule(dynamic,1)
    	\For {$kl$ = 1, ${kl}_{max}$}
    		\State $k,l\gets kl$ \Comment Deduce K and L indices
       		\State $screened\gets$ schwartz($i,j,k,l$) \Comment Schwartz screening
           		\If {\textbf{not} $screened$}
                	\State \textbf{call} eri($i,j,k,l,X_{ijkl}$) \Comment Calculate $(i,j|k,l)$
                    \LeftComment {Update private partial Fock matrices:}
                	\State $F_I(:,ithread)_{j,k,l}$+=$X_{ijkl}\cdot D_{kl,jl,jk}$ 	\label{alg:ompsh:fi}
                	\State $F_J(:,ithread)_{k,l}$+=$X_{ijkl}\cdot D_{il,ik}$
                    \LeftComment {Update shared Fock matrix:} 						\label{alg:ompsh:fj}
                	\State $Fock(k,l)$+=$X_{ijkl}\cdot D(i,j)$ 						\label{alg:ompsh:fk}
           		\EndIf
		\EndFor
    	\State !\$omp end do
		\State $Fock(:,j)$+=$\sum F_J(:,1$:$nthreads)$		\Comment Flush $F_J$	\label{alg:ompsh:iflush}
        \State !\$omp barrier
    	\State $i_{old} \gets i$													\label{alg:ompsh:iold}
    \EndIf

\EndLoop

\LeftComment Flush remainder $F_i$ contribution to $Fock$:
\State $Fock(:,i)$+=$\sum F_I(:,1$:$nthreads)$
\State !\$omp end parallel

\State \textbf{call} ddi\_gsumf($Fock$) \Comment 2e-Fock matrix reduction over MPI
\end{algorithmic}
\end{algorithm}

The access pattern of the Fock matrix by $k$ index corresponds to only one Fock matrix element. If threads have different $k$ and $l$ shell indices, it would be possible to skip saving data to the $k$ buffer and instead, to directly update the corresponding parts of the full Fock matrix. This condition will be satisfied if OpenMP parallelization over $k$ and $l$ loops is used. In this case, private storage is necessary for only the $i$ and $j$ blocks of the Fock matrix.

In the shared Fock matrix algorithm (\Cref{alg:ompsh}) the original four loops (\Cref{alg:mpi}) are arranged into two merged index loops. The first and second loops correspond to the combined $ij$ and $kl$ indices, respectively. MPI parallelization is executed over the top~($ij$) loop, while OpenMP parallelization is accomplished over the inner ($kl$) loop. In contrast to the private Fock matrix algorithm (\Cref{alg:omp}), this partitioning favors computer systems with a large number of MPI ranks and is the preferred strategy because this implementation of MPI iteration space is larger and the load balance is finer. By using this partitioning, it is also possible to utilize Schwarz screening across the $i$ and $j$ indices. Partitioning is especially important for very large jobs with very sparse ERI tensor because it allows the user to completely skip the most costly top-loop iterations.

Another difference from the private Fock matrix algorithm is that the ERI contribution is now added in three places (\Cref{alg:ompsh:fi}, lines~\ref{alg:ompsh:fi}-\ref{alg:ompsh:fk}): to the private $i$ buffer ($F_{ij}$, $F_{ik}$, $F_{il}$), the private $j$ buffer ($F_{jk}$, $F_{jl}$), and the shared Fock matrix ($F_{kl}$). At the end of the joint $kl$-loop, the partial Fock matrix contribution from $i$ and $j$ buffers needs to be added to the full Fock matrix. It is computationally expensive for a multithreaded environment because it requires explicit thread synchronization. However, it is possible to reduce the frequency of $i$ buffer flushing. After each $kl$ loop, the $i$ index very likely remains the same and there will be no need for $i$ buffer flushing. In the present algorithm, the old $i$ index is saved after the $kl$ loop (\Cref{alg:ompsh:fi}, line~\ref{alg:ompsh:iold}). The flushing of the $i$ buffer contribution to the Fock matrix is only done if the $i$ index were changed since the last iteration. Flushing the $j$ buffer is still required after each $kl$ loop (\Cref{alg:ompsh}, line~\ref{alg:ompsh:iflush}).

\begin{figure}
	\includegraphics[width=\columnwidth]{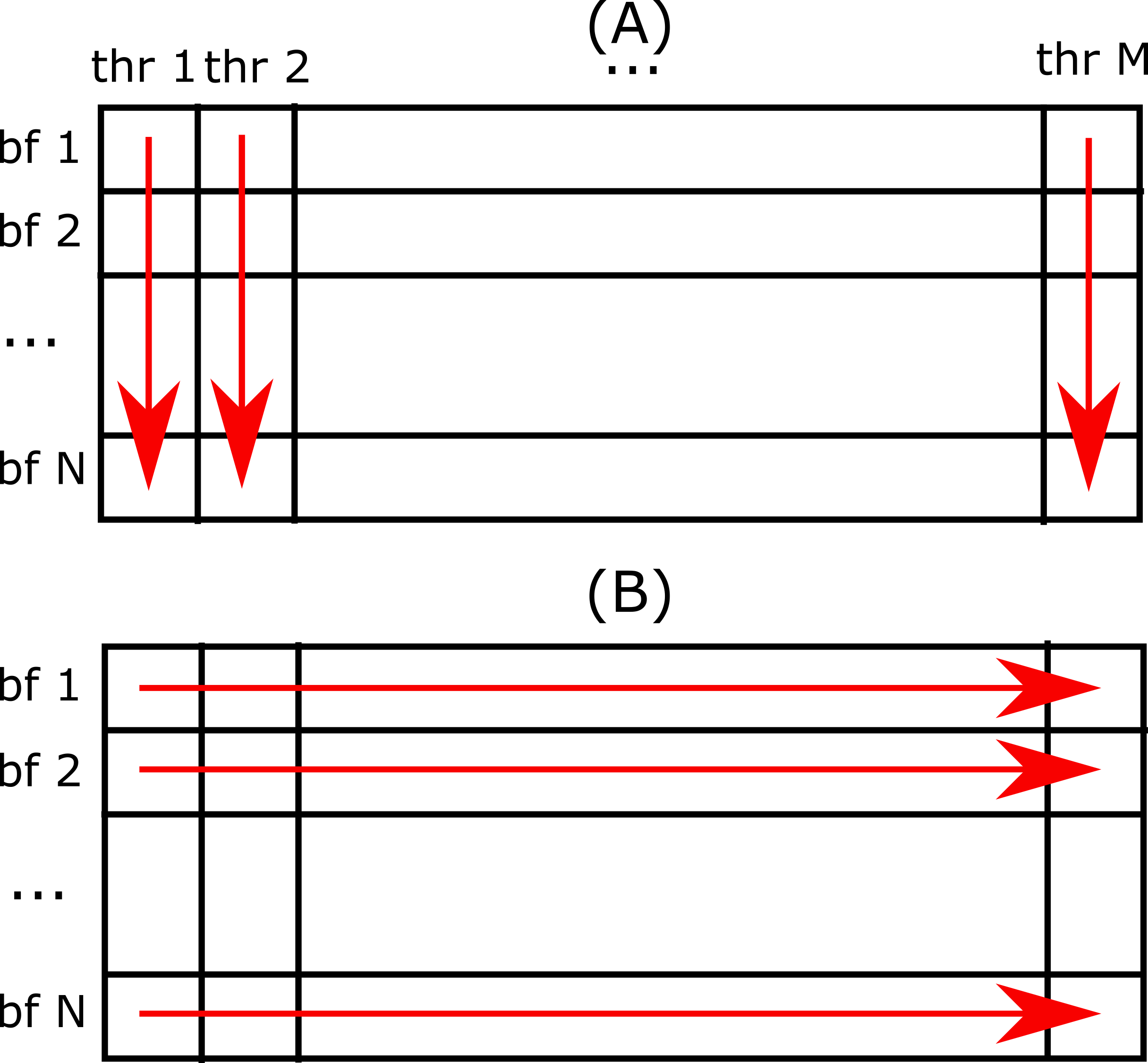}
	\caption{I and J Fock vectors update (A) and summing up all Fock elements for each Fock element in the vectors (B).
    		 ``bf'' means basis function and ``thr'' means thread.}
    \label{fig:buffer}
\end{figure}

A special array structure is needed for flushing and reducing buffers for the $i$ and $j$ blocks. Buffers are organized as two-dimensional arrays. The outer dimension of these arrays corresponds to threads, and the inner dimension corresponds to the data. Using Fortran notation, data is stored in matrix columns, with each thread displayed in its own column. This (column-wise) access pattern is used when threads add an ERI contribution to the buffers (\Cref{fig:buffer}~(A)). The access pattern is different when it is necessary to flush a buffer into the full Fock matrix. The tree-reduction algorithm is used to sum up the contribution from different columns and add them to the full Fock matrix. In this case, the access of threads to this matrix is row-wise (\Cref{fig:buffer}~(B)). Padding bytes were added to the leading dimension of the array and chunking was used on the reduction step to prevent false sharing. After the buffer is flushed into the Fock matrix, it is filled in with zeroes and is ready for the next cycle.

\section{Methodology}
\label{sec:benchmark}

\subsection{Description of hardware and software}
\label{ssec:supercomp}
The benchmarks reported in this paper were performed on the Intel Xeon Phi systems provided by the Joint Laboratory for System Evaluation (JLSE) and the Theta supercomputer at the Argonne Leadership Computing Facility (ALCF) \cite{alcf}, which is a part of the U.S. Department of Energy (DOE) Office of Science (SC) Innovative and Novel Computational Impact on Theory and Experiment (INCITE) program \cite{incite}. Theta is a 10-petaflop Cray XC40 supercomputer consisting of 3,624 Intel Xeon Phi 7230 processors. Hardware details for the JLSE and Theta system are shown in \Cref{tab:hw}.

The Intel Xeon Phi processor used in this paper has 64 cores each equipped with L1 cache. Each core also has two Vector Processing Units, both of which need to be used to get peak performance. This is possible because the core can execute two instructions per cycle. In practical terms, this can be achieved by using two threads per core. Pairs of cores constitute a tile. Each tile has an L2 cache symmetrically shared by the core pair. The L2 caches between tiles are connected by a two dimensional mesh. The cores themselves operate at 1.3 GHz. Beyond the L1 and L2 cache structure, all the cores in the Intel Xeon Phi processor share 16 GBytes of MCDRAM (also known as High Bandwidth Memory) and 192 GBytes of DDR4. The bandwidth of MCDRAM is approximately 400 GBytes/sec while the bandwidth of DDR4 is approximately 100 GBytes/sec. 

\begin{table}
  \caption{Hardware and software specifications}
  \label{tab:hw}

  \begin{tabularx}{\columnwidth}{XX}
  \toprule
			\multicolumn{2}{c}{\textbf{\intelphi\ node characteristics}} \\
    \midrule 
    \intelphi\ models				&	7210 and 7230 (64~cores, 1.3~GHz, 
    									2,622 GFLOPs) \\
    Memory per node					&	16 GB MCDRAM, \newline 192 GB DDR4 RAM \\
    Compiler						&	Intel Parallel Studio XE 2016v3 \\
    \midrule
    		\multicolumn{2}{c}{\textbf{JLSE \iphi\ cluster (26.2 TFLOPS peak)}} \\
    \midrule
    \# of \intelphi\ nodes	&	10 \\
    Interconnect type				&	Intel Omni-Path\textsuperscript{TM} \\
    \midrule
    		\multicolumn{2}{c}{\textbf{Theta supercomputer (9.65~PFLOPS peak)}} \\
    \midrule
    \# of \intelphi\ nodes				&	3,624 \\
    Interconnect type				&	Aries interconnect with \newline Dragonfly topology \\
  \bottomrule
\end{tabularx}

\end{table}

\begin{table}
\begin{threeparttable}
  \caption{Chemical systems used in benchmarks and their size characteristics}
  \label{tab:chem}

  \begin{tabularx}{\columnwidth}{XYYYYY}

  \toprule

  \multirow{2}{*}{Name}	&	\multirow{2}{*}{\# atoms}	&	\multirow{2}{*}{\# BFs\tnote{a}}	&	\multicolumn{3}{c}{Memory footprint\tnote{b}, GB} \\
  \cmidrule(l){4-6}
  		& & &	{MPI\tnote{c}}	&	{Pr.F.\tnote{d}}	&	{Sh.F.\tnote{e}} \\
  \midrule
  	0.5~nm	&	44			&	660			&	7		&	0.13		&	0.03	\\
	1.0~nm	&	120			&	1800		&	48		&	1			&	0.2	\\
	1.5~nm	&	220			&	3300		&	160		&	3			&	0.8	\\
	2.0~nm	&	356			&	5340		&	417		&	8			&	2	\\
	5.0~nm	&	2016		&	30240		&	9869	&	257			&	52	\\
	\bottomrule
  \end{tabularx}

  \begin{tablenotes}
  	\item [a] BF -- basis function
    \item [b] Estimated using \crefrange{eqn:mem:mpi}{eqn:mem:shr}
 	\item [c] MPI-only SCF code
    \item [d] Private Fock SCF code
    \item [e] Shared Fock SCF code
  \end{tablenotes}
\end{threeparttable}
\end{table}

These two levels of memory can be configured in three different ways (or modes). The modes are referred to as Flat mode, Cache mode, and Hybrid mode. Flat mode treats the two levels of memory as separate entities. The Cache mode treats the MCDRAM as a direct mapped L3 cache to the DDR4 layer. Hybrid mode allows the user to use a fraction of MCDRM as L3 cache allocate the rest of the MCDRAM as part of the DDR4 memory.
In Flat mode, one may choose to run entirely in MCDRAM or entirely in DDR4. The "numactl" utility provides an easy mechanism to select which memory is used. It is also possible to choose the kind of memory used via the "memkind" API, though as expected this requires changes to the source code.

Beyond memory modes, the Intel Xeon Phi processor supports five cluster modes. The motivation for these modes can be understood in the following manner: to maintain cache coherency the Intel Xeon Phi processor employs a distributed tag directory (DTD). This is organized as a set of per-tile tag directories (TDs), which identify the state and the location on the chip of any cache line. For any memory address, the hardware can identify the TD responsible for that address. The most extreme case of a cache miss requires retrieving data from main memory (via a memory controller). It is therefore of interest to have the TD as close as possible to the memory controller. This leads to a concept of locality of the TD and the memory controllers.
It is in the developer's interest to maintain the locality of these messages to achieve the lowest latency and greatest bandwidth of communication with caches. Intel Xeon Phi supports all-to-all, quadrant/hemisphere and sub-NUMA cluster SNC-4/SNC-2 modes of cache operation.

For large problem sizes, different memory and clustering modes were observed to have little impact on the time to solution for the three versions of the GAMESS code. For this reason, we simply chose the mode most easily available to us. In other words, since the choice of mode made little difference in performance, our choice of Quad-Cache mode was ultimately driven by convenience (this being the default choice in our particular environment). Our comments here apply to large problem sizes, so for small problem sizes, the user will have to experiment to find the most suitable mode(s).

\subsection{Description of chemical systems}
\label{ssec:chemical}
\begin{figure}
	\includegraphics[width=\columnwidth]{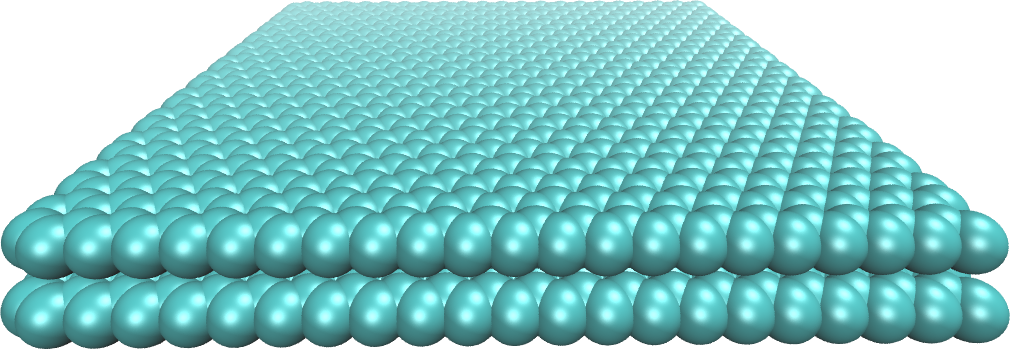}
	\caption{Model system of a C$_{2016}$ graphene bilayer. In the text, we refer to this system as 5~nm.
    		 There are two layers with size 5~nm by 5~nm.
             Each graphene layer consists of 1,008 carbon atoms.}
    \label{fig:graphene}
\end{figure}

For benchmarks, a system consisting of parallel series of graphene sheets was chosen. This system is of interest to researchers in the area of (micro)lubricants \cite{kawai2016superlubricity}. A physical depiction of the configuration is provided in \Cref{fig:graphene}. The graphene-sheet system is ideal for benchmarking, because the size of the system is easily manipulated. Various Fock matrix sizes can be targeted by adjusting the system size.

\subsection{Characteristics of datasets}
\label{ssec:datasets}
In all, five configurations of the graphene sheets system were studied. The datasets for the systems studied are labeled as follows: 0.5~nm, 1.0~nm, 1.5~nm, 2.0~nm, and 5.0~nm.  \Cref{tab:chem} lists size characteristics of these configurations. The same 6-31G(d) basis set (per atom) was used in all calculations. For N basis functions, the density, Fock, AO overlap, one-electron Fock matrices and the matrix of MO coefficients are N$\times$N in size. These are the main data structures of significant size. Therefore, the benchmarks performed in this work process matrices which range from 660$\times$660 to 30,240$\times$30,240. For each of the systems studied, \Cref{tab:chem} lists the memory requirements of the three versions of GAMESS HF code.
Denoting $N_{BF}$ as the number of basis functions, the following equations describe the asymptotic $(N_{BF}\to\infty)$ memory footprint for the studied HF algorithms:
\begin{subequations}
	\label{eqn:mem}
	\begin{align}
		M_{MPI} =& 5/2 \cdot N_{BF}^2 \cdot N_{MPI\_per\_node}, 				\label{eqn:mem:mpi} \\
		M_{PrF} =& (2+N_{threads}) \cdot N_{BF}^2 \cdot N_{MPI\_per\_node}, 	\label{eqn:mem:prv} \\
		M_{ShF} =& 7/2 \cdot N_{BF}^2 \cdot N_{MPI\_per\_node},					\label{eqn:mem:shr}
	\end{align}
\end{subequations}
where $M_{MPI}$, $M_{PrF}$, $M_{ShF}$ denote the memory footprint of MPI-only, private Fock, and shared Fock algorithms respectively; $N_{threads}$ denotes the number of threads per MPI process for the OpenMP code, and $N_{MPI\_per\_node}$ denotes the number of MPI processes per KNL node. For OpenMP runs $N_{MPI\_per\_node}=4$, while for MPI runs the number of MPI ranks was varied from 64 to 256.

If one compares columns MPI versus Pr.F and Sh.F. in \Cref{tab:chem}, you will see that the private Fock code has about a 50 times less footprint compared to the stock MPI code. For the shared Fock code, the difference is even more dramatic with a savings of about 200 times. The ideal difference is 256 times since we compare 256 MPI ranks in the stock MPI code where all data structures are replicated versus 1 MPI rank with 256 threads for the hybrid MPI/OpenMP codes. But we introduced additional replicated structures (see \Cref{fig:buffer}) and many relatively small data structures are replicated also in the MPI/OpenMP codes. This explains the difference between the ideal and observed footprints.

Each of the aforementioned datasets was used to benchmark three versions of the GAMESS code. The first version is the stock GAMESS MPI-only release that is freely available on the GAMESS website~\cite{gamesswebsite}. The second version is a hybrid MPI/OpenMP code, derived from the stock release. This version has a shared density matrix, but a thread-private Fock matrix. The third version of the code is in turn derived from the second version; it has shared density and Fock matrices. A key objective was to see how these incremental changes allow one to manage (i.e., reduce) the memory footprint of the original code while simultaneously driving higher performance.

\section{Results}
\label{sec:results}

\subsection{Single node performance}
\label{ssec:singlenode}
\begin{figure}
	\includegraphics[width=\columnwidth]{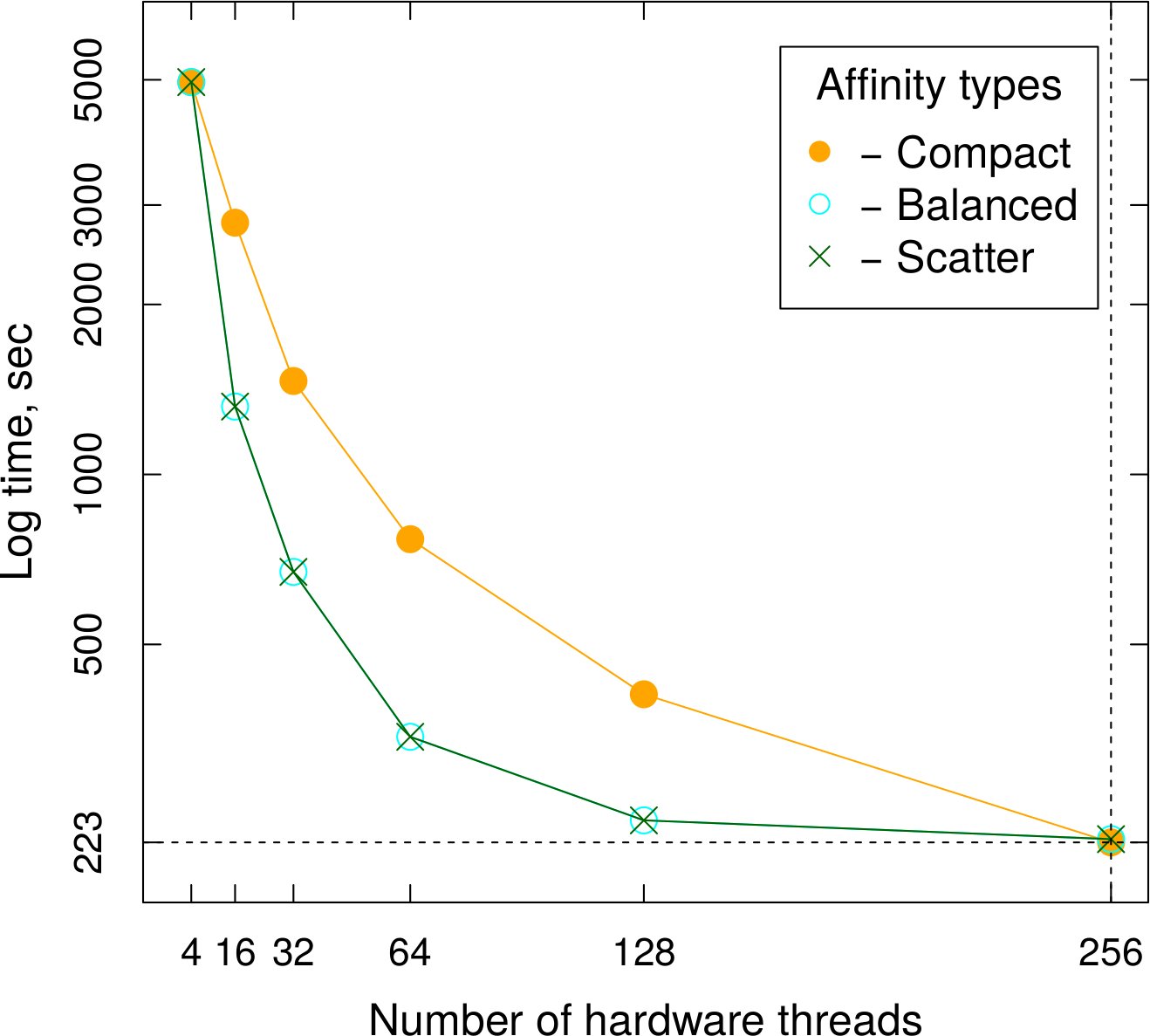}
	\caption{Performance dependence on OpenMP thread affinity type for the shared Fock version of the GAMESS code
    		 on a single \intelphireg\ processor using the 1.0 nm benchmark.
             All calculations are performed in quad-cache mode.
             Four MPI ranks were used in all cases.
             The number of threads per MPI rank was varied from 1 to 64.}
    \label{fig:afty}
\end{figure}

The second generation Intel Xeon Phi processor supports four hardware threads per physical core. Generally, more threads per core can help hide latencies inherent in an application. For example, when one thread is waiting for memory, another can use the processor. The out-of-order execution engine is beneficial in this regard as well. To manipulate the placement of processes and threads, the \verb|I_MPI_DOMAIN| and \verb|KMP_AFFINITY| environment variables were used. 
We examined the performance picture when one thread per core is utilized and when four threads per core are utilized. As expected, the benefit is highest for all versions of GAMESS for two threads (or processes) per core. For three and four threads per core, some gain is observed, albeit at a diminished level. \Cref{fig:afty} shows the scaling curves with respect to the number of hardware threads utilized observed by us.

\begin{figure}
	\includegraphics[width=\columnwidth]{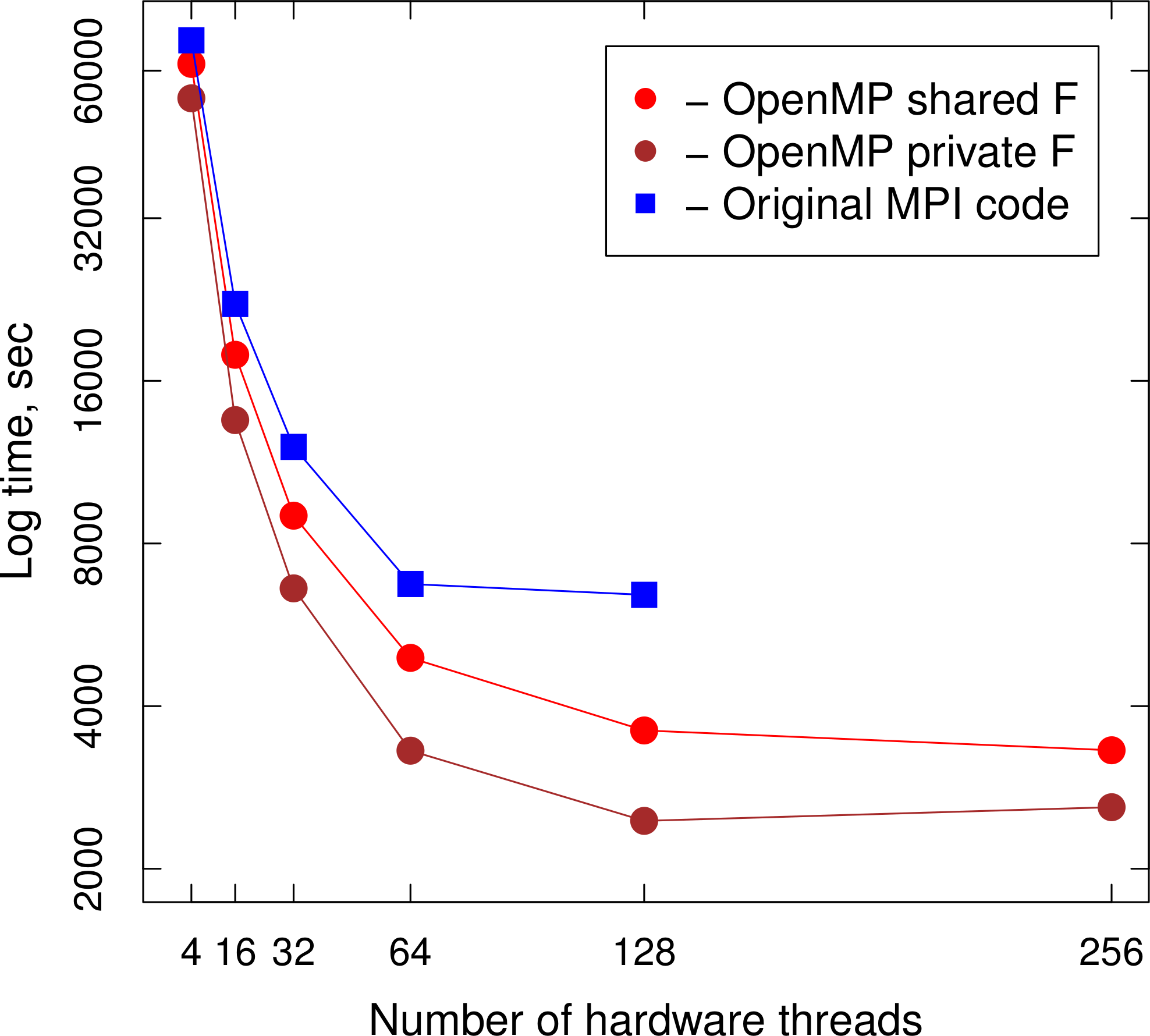}
	\caption{Scalability with respect to the number of hardware threads of the original MPI code
    		and two OpenMP versions on a single \intelphireg\ processor using the 1.0~nm benchmark.}
    \label{fig:singlescaling}
\end{figure}

As a first test, single-node scalability was examined with respect to hardware threads of all three versions of GAMESS. For the MPI-only version of GAMESS, the number of ranks was varied from~4 to~256. For the hybrid versions of GAMESS, the number of ranks times the number of threads per rank is the number of hardware threads targeted. The larger memory requirements of the original MPI-only code restrict the computations to, at most, 128 hardware threads. In contrast, the two hybrid versions can easily utilize all 256 hardware threads available. Finally, in general terms, on cache based memory architectures, it is expected that larger memory footprints potentially lead to more cache capacity and cache line conflict effects. These effects can lead to diminished performance, and this is yet another motivation to look at a hybrid MPI+X approach.

The results of our single-node tests are plotted in \Cref{fig:singlescaling}. It is found that using the private Fock version leads to the best time to solution for the 1.0~nm dataset, for any number of hardware threads. This version of the code is much more memory-efficient than the stock version but, because the Fock matrix data structure is private, it has a much larger memory footprint than the shared Fock version of GAMESS. Nevertheless, because the Fock matrix is private, there is less thread contention than the shared Fock version.

It was mentioned in \Cref{ssec:omp} that shared Fock algorithm introduces additional overhead for thread synchronization. For small numbers of Intel Xeon Phi threads, this overhead is expected to be low. Therefore the shared Fock version is expected to be on par with the other versions. Eventually, as the overhead of the synchronization mechanisms begins to increase, the private Fock version of the code is found to dominate. In the end, the private Fock version outperforms stock GAMESS because of the reduced memory footprint, and outperforms the shared Fock version because of a lower synchronization overhead.
Therefore, on a single node, the private Fock version gives the best time-to-solution of the three codes, but the shared Fock version strikes a (better) balance between memory utilization and performance.

\begin{figure}
	\includegraphics[width=\columnwidth]{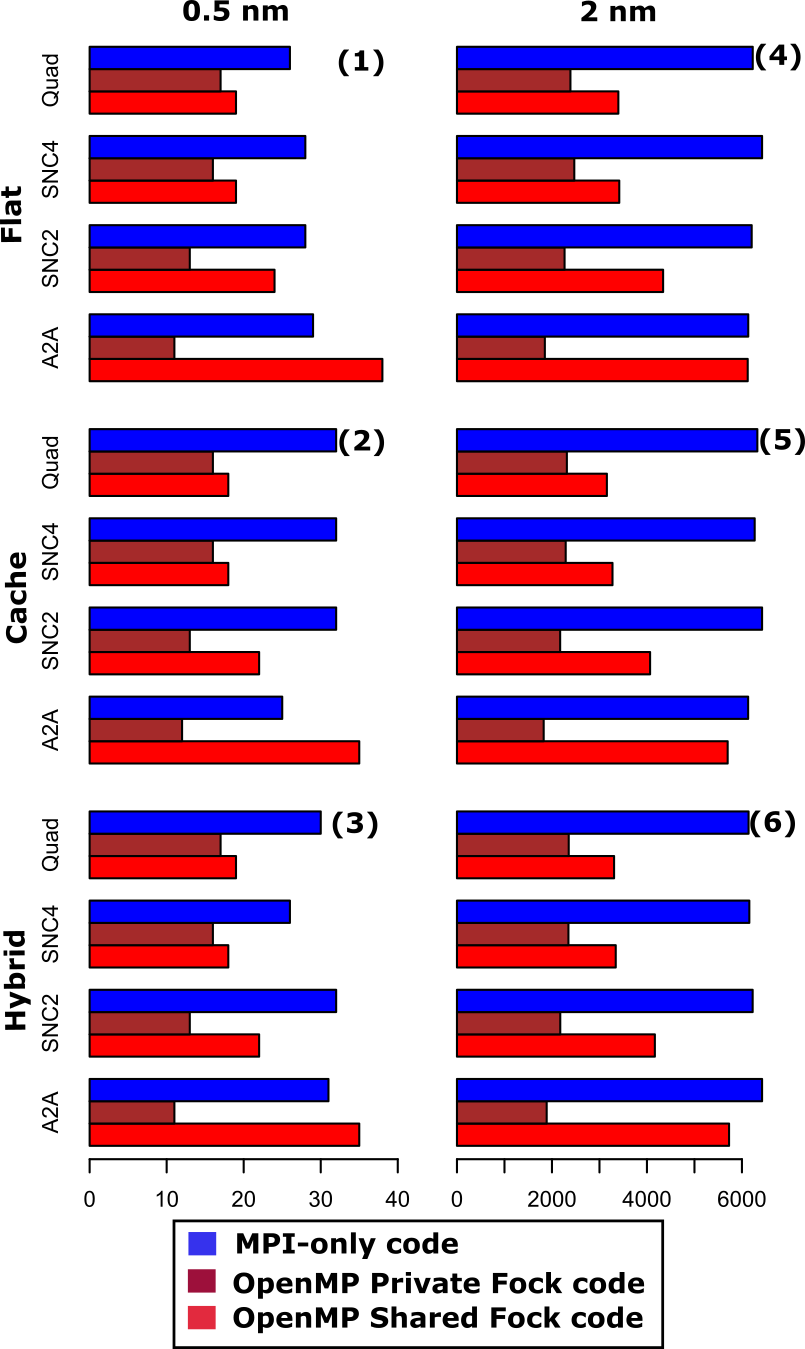}
	\caption{Time to solution (x axis, time in seconds) for different clustering and memory modes.
    		 Left column displays the small chemical system -- 0.5~nm bilayer graphene and
             right column displays one of the largest molecules bilayer graphene -- 2.0~nm.}
    \label{fig:tts}
\end{figure}

Beyond this, one must consider the choice of memory mode and cluster mode of the Intel Xeon Phi processor. It should be noted that, depending on the compute and memory access patterns of a code, the choice of memory and cluster mode can be a potentially significant performance variable. The performance impact of different memory and cluster modes is examined for the 0.5~nm (small) and~2.0~nm (large) datasets. The results are shown in \Cref{fig:tts}. For both datasets, some variation in performance is apparent when different cluster modes and memory modes are used. The smaller dataset indicates more sensitivity to these variables than the larger dataset. Also, for both data sizes the private Fock version performs best in all cluster and memory modes tested. Also, except in the All-to-All cluster mode, the shared Fock version significantly outperforms the MPI-only stock version. In the All-to-All mode, the MPI-only version actually outperforms the shared Fock version for small datasets, and the two versions are close to parity for large datasets. In total, it is concluded that the quadrant-cache cluster-memory mode is best suited to the design of the GAMESS hybrid codes.

\subsection{Multi-node performance}
It is very important to note that the total number of MPI ranks for GAMESS is actually twice the number of compute ranks because of the DDI. The DDI layer was originally implemented to support one-sided communication using MPI-1. For GAMESS developers, the benefit of DDI is convenience in programming. The downside is that each MPI compute process is complemented by an MPI data server~(DDI) process, which clearly results in increased memory requirements. Because data structures are replicated on a rank-by-rank basis, the impact of DDI on memory requirements is particularly unfavorable to the original version of the GAMESS code. To alleviate some of the limitations of the original implementation, an implementation of DDI based on MPI-3 was developed \cite{pruitt2016private}. Indeed, by leveraging the ``native'' support of one-sided communication in MPI-3, the need for a DDI process alongside each MPI rank was eliminated. For all three versions of the code benchmarked here, no DDI processes were needed.

\begin{figure}
	\includegraphics[width=\columnwidth]{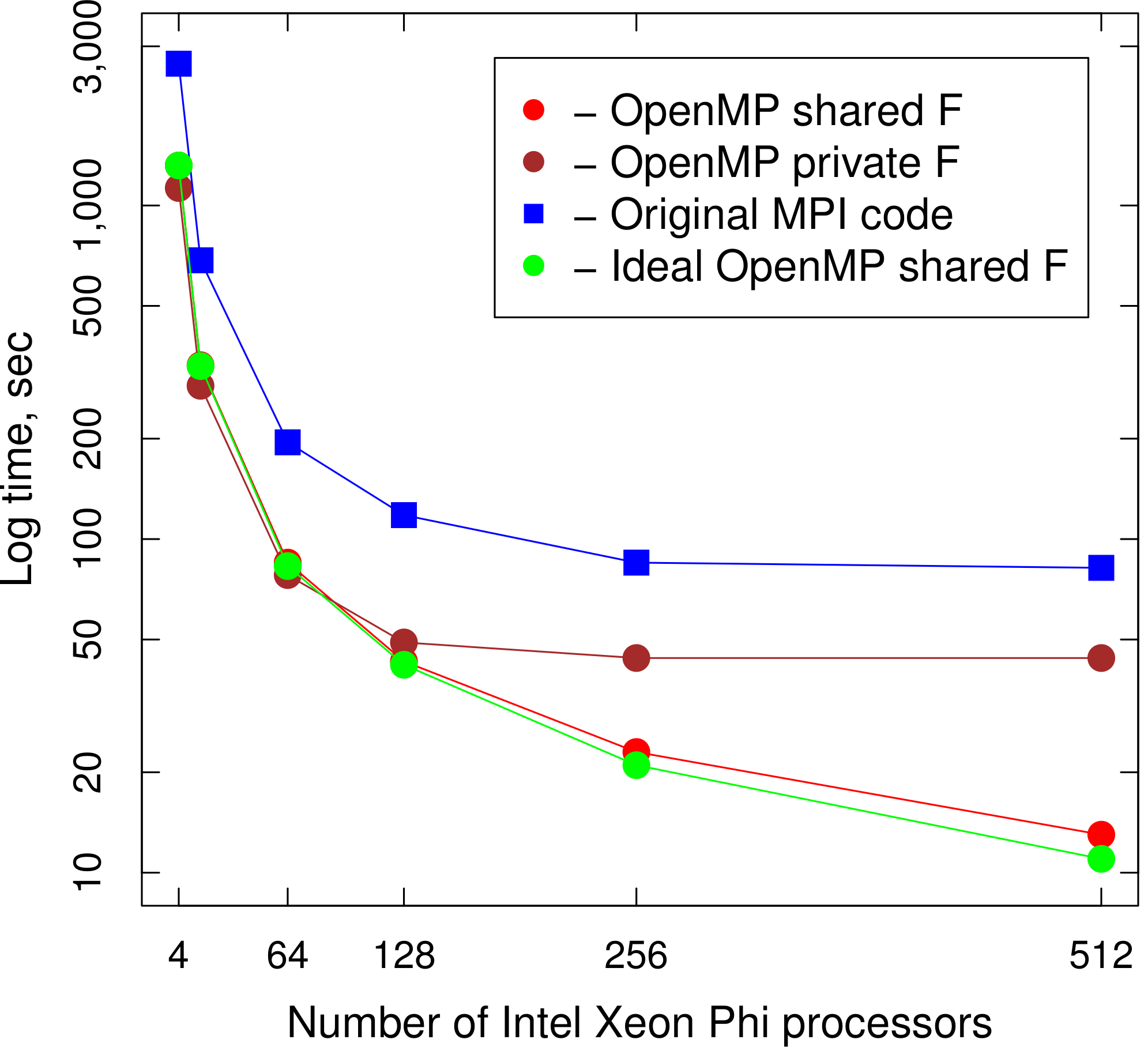}
	\caption{Multi-node scalability of the Private Fock and the Shared Fock hybrid MPI-OpenMP
    		 and the MPI-only stock GAMESS codes on the Theta machine with the 2.0~nm dataset.
             The quad-cache cluster-memory mode was used for all data points.}
    \label{fig:2nm}
\end{figure}

\Cref{fig:2nm} shows the multi-node scalability of the MPI-only version of GAMESS versus the private Fock and the shared Fock hybrid versions. It is important to appreciate at the outset that the multi-node scalability of the original MPI-only version of GAMESS is already reasonable. For example, the code scales linearly to 256 Xeon Phi nodes, and it is really the memory footprint bottleneck that limits how well all the Xeon Phi cores on any given node can be used. This pressure is reduced in the private Fock version of the code, and it is essentially eliminated in the shared Fock version. Overall, for the 2~nm dataset, the shared Fock code runs about six times faster than stock GAMESS on 512 Xeon Phi processors. It resulted from the better load balance of the shared Fock algorithm that uses all four shell indices -- two are used in MPI and two are used in OpenMP workload distribution. The actual timings and efficiencies are listed in \Cref{tab:efficiency}.

\begin{table}
\begin{threeparttable}
  \caption{Parallel efficiency of the three different HF algorithms using 2.0~nm dataset}
  \label{tab:efficiency}
  \begin{tabularx}{\columnwidth}{XYYYYYY}

  \toprule
    	\multirow{2}{*}{\# Nodes}		&	\multicolumn{3}{c}{Time-to-solution, s} &
                            \multicolumn{3}{c}{Parallel efficiency, \%} \\
        \cmidrule(rl{0.75em}){2-4} \cmidrule(l){5-7}
  					&	{MPI\tnote{a}} &	{Pr.F.\tnote{a}} &	{Sh.F.\tnote{a}} &
                        {MPI\tnote{a}} &	{Pr.F.\tnote{a}} &	{Sh.F.\tnote{a}} \\

  	\midrule
		4	&	2661	&	1128	&	1318	&	100	&	100	&	100 \\
		16	&	685		&	288		&	332		&	97	&	98	&	99 \\
		64	&	195		&	78		&	85		&	85	&	90	&	97 \\
		128	&	118		&	49		&	43		&	70	&	72	&	96 \\
		256	&	85		&	44		&	23		&	49	&	40	&	90 \\
		512	&	82		&	44		&	13		&	25	&	20	&	79 \\
    \bottomrule
   \end{tabularx}

 	\begin{tablenotes}
 		\item [a] MPI-only SCF code
    	\item [b] Private Fock SCF code
    	\item [c] Shared Fock SCF code
 	\end{tablenotes}
\end{threeparttable}
\end{table}

\begin{figure}
	\includegraphics[width=\columnwidth]{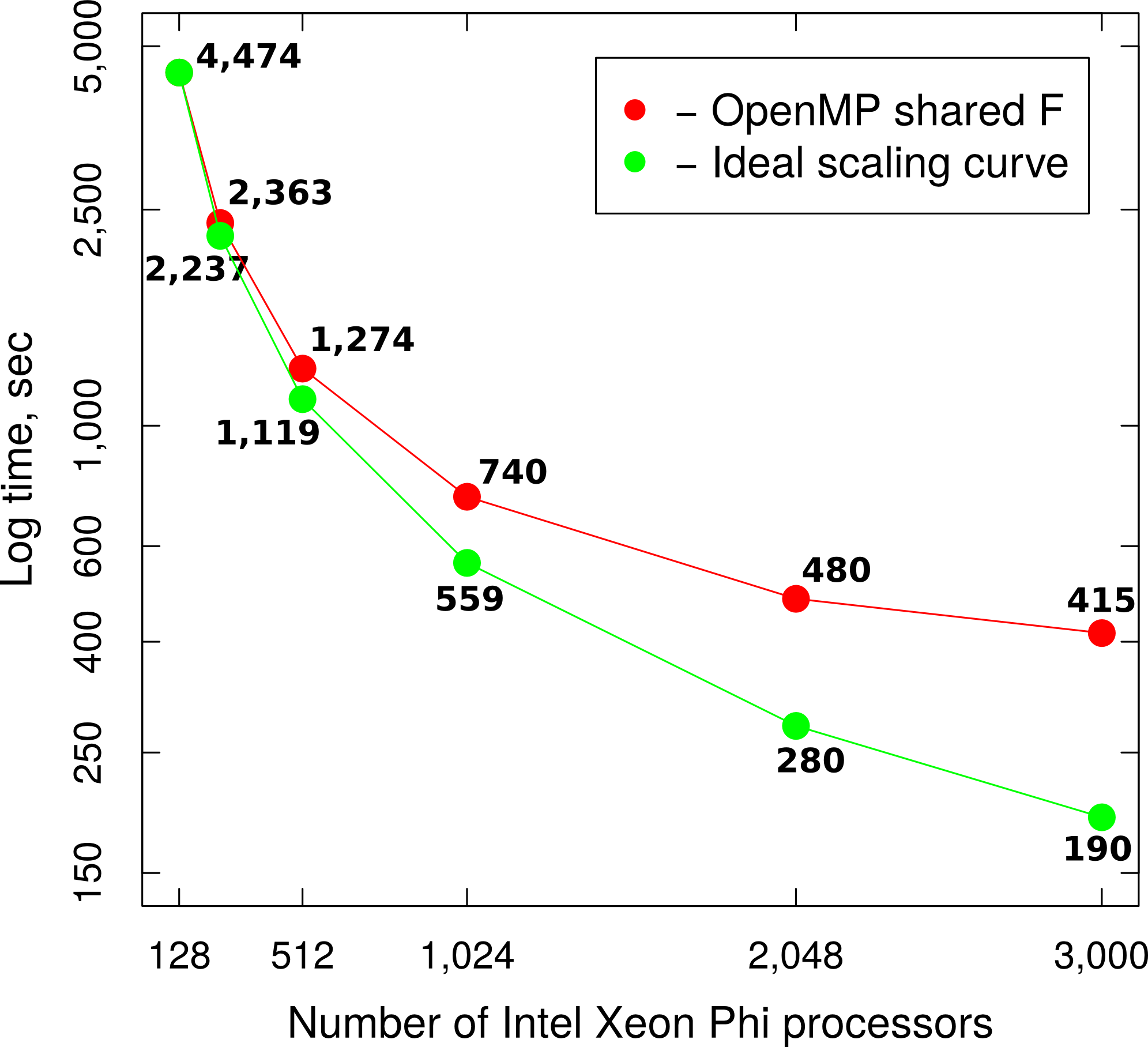}
	\caption{Scalability of the Shared Fock hybrid MPI-OpenMP version of GAMESS on the Theta machine
    		 for the 5.0~nm (i.e. large) dataset in quadrant cache mode on 3,000 \intelphireg\ processors.
             The results here are for 4~MPI ranks per node with 64~threads per rank,
             giving full saturation (in terms of hardware threads) on every \intelphireg\ node. For each point in the figure, we show the time in seconds.}
    \label{fig:5nm}
\end{figure}

\Cref{fig:5nm} shows the behavior of the shared Fock version of GAMESS for the 5~nm dataset. It is the largest dataset we could fit in memory on Theta. Since we run on 4~MPI ranks the memory footprint is approximately 208~GB per node. This figure shows good scaling of the code up to 3,000 Xeon Phi nodes, which is equal to 192,000 cores (64~cores per node).

\section{Conclusion}
\label{sec:conclusion}

In this paper, conversion of the MPI-only GAMESS HF code to hybrid MPI-OpenMP versions is described. The resulting hybrid implementations are benchmarked to exhibit improvements in the time-to-solution and memory footprint compared to the original MPI-only version. The code design decisions taken here were justified and implemented in a systematic way. Focus was placed on sharing the two primary (memory consuming) objects, the density and Fock matrices, in the SCF loop among the computation units. To the best of our knowledge, having a shared Fock matrix is an unique feature of our implementation. Indeed, this is absent in all other threaded HF codes known to us.

We have discussed two new HF implementations, each of which maintains full functionality of the underlying GAMESS code. In the first version, the density matrix was shared across threads, while the Fock matrix was kept private. The second version leveraged the first step, and focused entirely on making the Fock matrix a shared object. As a result, the memory footprint of the original code was lowered systematically while improving cache utilization and time-to-solution. Clearly, we have taken only the first steps towards an efficient hybrid HF implementation in GAMESS. In future work, we plan to tune our hybrid OpenMP/MPI code more thoroughly.

Our new hybrid MPI/OpenMP codes significantly outperform the official stock MPI-only code in GAMESS. Our best case implementation has about 200 times smaller memory footprint and runs up to 6 times faster than the original MPI-only version. Both our hybrid versions also have better scalability with respect to cores and nodes on single node and multi-node Intel Xeon Phi systems respectively.

It is also noted that the code optimizations reported in this paper are expected to be applicable to all previous and future generations of Intel Xeon Phi processors, as well as beneficial on the Intel Xeon multicore platform. The fact that the code already scales well on a large number of second generation Intel Xeon Phi processors enables us to help bring the promise of the ``many-core'' philosophy to the large scientific community that has long benefited from the extensive functionality of the GAMESS code. Like the MPI-only version, the hybrid versions of GAMESS can be deployed on systems ranging from a single desktop to large supercomputers. In addition, the hybrid codes offer enhanced configurability and parallel granularity.

Finally, the lessons learned here are applicable to virtually any code that handles non-linear partial differential equations using a matrix representation. In this paper, we treat the problem of assembling a matrix in parallel subject to highly non-regular data dependencies. Indeed, a variety of methods, such as Unrestricted Hartree Fock (UHF), Generalized Valence Bond (GVB), Density Functional Theory (DFT), and Coupled Perturbed Hartree-Fock (CPHF), all have this structure. The implementation of these methods can therefore directly benefit from this work. Beyond quantum chemistry, we note, the SCF approach shares much in common  with generic non-linear solvers. We therefore conclude that the strategies discussed in this work are directly applicable to computer programs encountered in other areas of science.

\begin{acks}
  This research used the resources of the Argonne Leadership Computing Facility, which is a U.S. Department of Energy (DOE) Office of Science User Facility supported under Contract DE-AC02-06CH11357. We gratefully acknowledge the computing resources provided and operated by the Joint Laboratory for System Evaluation (JLSE) at Argonne National Laboratory. Dr. Kristopher Keipert and Prof. Mark S. Gordon acknowledge the support of a Department of Energy Exascale Computing Project grant to the Ames Laboratory. We thank the \grantsponsor{}{{\intelreg} Parallel Computing Centers program}{} for funding. The authors would like to thank the RSC Technologies staff for the discussions and help.

\end{acks}

\balance

\bibliographystyle{ACM-Reference-Format}
\bibliography{sc2017}

\newpage

\appendix
\section{Artifact Description: An efficient MPI/OpenMP parallelization of the Hartree-Fock method for the second generation of \intelphireg\ processor}

\subsection{Abstract}

This description contains all the information needed to run the simulations described in the SC17 paper ``An efficient MPI/OpenMP parallelization of the Hartree-Fock method for the second generation of \intelphireg\ processor''. More precisely, we explain how to compile and run GAMESS simulations to reproduce results presented in \Cref{sec:results} of the paper. The PDB files, GAMESS inputs, and submission scripts used in this work can be downloaded from the Git repository available online\footnote{\label{gitrepo}\url{https://github.com/gms-bbg/gamess-papers}, folder Mironov\_Vladimir-2017-SC17}.

\subsection{Description}

\subsubsection{Check-list (artifact meta information)}

{\em Fill in whatever is applicable with some informal keywords and remove the rest}

{\small
\begin{itemize}
  \item {\bf Algorithm: } Self Consistent Field (SCF) algorithm in Hartree-Fock (HF) method
  \item {\bf Program: } GAMESS binary, Fortran and C libraries
  \item {\bf Compilation: } \intelreg\ Parallel Studio XE 2016 v3.210
  \item {\bf Data set: } graphene bilayer systems from 0.5 nm to 5 nm; details are in the
  		main paper \cref{sec:benchmark} and in \Cref{tab:ad:chem}
  \item {\bf Run-time environment: } when running on Cray XC40 the following modules were loaded:
  		\begin{itemize}
        	\item craype/2.5.9
            \item PrgEnv-intel/6.0.3
            \item craype-mic-knl  
        \end{itemize}
  \item {\bf Hardware: } all single node benchmarked were run on JLSE cluster on \intelphireg\ 7210 processor; all multi-node benchmarks were run on Cray XC40 with up to 3,000 \intelphireg\ 7230 processors
  \item {\bf Experiment customization: } varied number of MPI ranks from 1 to 12,000 and number of OpenMP threads from 1 to 256 on single node.
  \item {\bf Publicly available?: } yes (partially)
\end{itemize}
}

\subsubsection{How software can be obtained (if available)}

In this work, we used GAMESS version dated August 18, 2016 Revision 1. The original GAMESS code can be downloaded (at no cost) from the official GAMESS website (registration required)\footnote{\url{http://www.msg.ameslab.gov/gamess/download.html}}.

Patches for GAMESS developed in this work can be obtained from the authors by request.

\subsubsection{Hardware dependencies}
All measurements in this work were performed on the \second\ generation of \intelphireg\ processor. However, the same benchmarks can be run on any other architectures.

\subsubsection{Software dependencies}
C (C99 standard) compiler, Fortran 95 compatible compiler with OpenMP 3.0 support, MPI library. The code has been extensively tested only for \intelreg\ Parallel Studio XE 2016 update 3 compilers and libraries. MKL BLAS library was used to link GAMESS, but it does not affect the performance of the SCF code and thus, it is optional.

\subsubsection{Datasets}
All structures of chemical systems and corresponding input files used for benchmarking code can be downloaded from the Git repository\cref{gitrepo}. They are easily reconfigurable bi-layer graphene systems. The problem size (computation time, memory footprint) depends on the number of basis functions for the system. These numbers are provided in \Cref{tab:ad:chem}. The basis set used in all calculations is 6-31G(d).

\begin{table}[ht]
	\centering
  	\caption{Chemical systems used in benchmarks and their size characteristics. BF stands for basis function.}
  	\label{tab:ad:chem}
  	\begin{tabular}{lccc}
    \toprule
    Name & \# atoms & \# shells & \# basis functions\\		 
    \midrule
    0.5 nm	&	44		&	176		&	660 \\
    1 nm	&	120		&	480		&	1,800 \\
	1.5 nm	&	220		&	880		&	3,300 \\
	2 nm	&	356		&	1,424	&	5,340 \\
	5 nm	&	2,016	&	8,064	&	30,240 \\
  	\bottomrule
\end{tabular}
\end{table}

\subsection{Installation}

We followed the standard installation procedure outlined in the file \texttt{PROG.DOC} in GAMESS root directory. GAMESS uses a custom build configuration system. The first step is configuration of the \texttt{install.info} file. To perform the basic configuration one need to run \texttt{\$\{GMS\_HOME\}/config} script and specify compilers and libraries for the compilation. When the script asks whether to compile GAMESS with LIBCCHEM one need to refuse. After that the \texttt{install.info} file can be edited directly to get the desired configuration. After setup is finished, GAMESS compilation can be done with \texttt{make} command. At successful conclusion, the file \texttt{gamess.\$(VERNO).x} will appear in GAMESS home directory, where \texttt{\$(VERNO)} is a variable specified at basic configuration step with \texttt{\$\{GMS\_HOME\}/config} script. \texttt{VERNO="00"} by default.

We used two different \intelphireg systems: JLSE and XC40. The actual \texttt{install.info} configurations are accessible at Git repository in folders \texttt{JLSE} and \texttt{CRAYXC40}.  The key parameters of the \texttt{install.info} file for both clusters are summarized in \Cref{tab:ad:clust}.

Moreover, we manually added the flag \texttt{-xMIC-AVX512} to the compilation line in \texttt{\$\{GMS\_HOME\}/comp} script and added \texttt{-mkl} flag to the link line in \texttt{\$\{GMS\_HOME\}/lked} script. On Cray XC40 system we also modified \texttt{\$\{GMS\_HOME\}/comp}, \texttt{\$\{GMS\_HOME\}/compall}, \texttt{\$\{GMS\_HOME\}/lked}, and \texttt{\$\{GMS\_HOME\}/ddi/compddi} scripts to add a new target ``cray-xc''. The modified files are accessible at Git repository in folders \texttt{JLSE} and \texttt{CRAYXC40}.

For DDI library, we used an experimental version of software to run all benchmarks for the single node \intelphireg\ performance on JLSE cluster. This DDI library uses one-sided communication features of MPI-3 which does not spawn data servers. On the Cray XC40 system, we used the standard DDI library.

\subsection{Experiment workflow}

We used the standard workflow of the experiment: compile code and run it for different \intelphireg\ system. We varied number of MPI ranks from 1 to 3,000, number of threads from 1 to 256 per rank, and varied different \iphireg\ clustering and memory modes.

\subsection{Evaluation and expected result}

We ran GAMESS on the Joint Laboratory for System Evaluation (JLSE) cluster using this submission line for OpenMP code:
\begin{lstlisting}[language=bash]
mpirun -n $2 \
  -env OMP_NUM_THREADS $3 \
  -env I_MPI_SHM_LMT shm \
  -env KMP_STACKSIZE 200m \
  -env I_MPI_PIN_DOMAIN auto \
  -env KMP_AFFINITY verbose \
  -env <GAMESS options> gamess.00.x
\end{lstlisting}

On the XC40 system, we used the following submission line using Cobalt queuing system for OpenMP code:

\begin{lstlisting}[language=bash]
rpn=4
allranks=$((COBALT_JOBSIZE*rpn))
aprun -n $allranks -N $rpn \
  -d $threads -cc depth -j 4 \
  --env OMP_NUM_THREADS=64 \
  --env I_MPI_PIN_DOMAIN=64:compact \
  --env KMP_STACKSIZE=2000m \
  --env <GAMESS options> gamess.00.x
\end{lstlisting}

For MPI code, we used this submission line:
\begin{lstlisting}[language=bash]
rpn=4
allranks=$((COBALT_JOBSIZE*rpn))
aprun -n $allranks -N $rpn \
  -cc depth -j 4 \
  --env I_MPI_PIN_DOMAIN=64:compact \
  --env <GAMESS options> gamess.00.x
\end{lstlisting}

The submission and run scripts are accessible at Git repository in folders \texttt{JLSE} and \texttt{CRAYXC40}.

The results of the computation are printed to \texttt{STDOUT} or redirected to the file specified in submission/run script. Time for the SCF part of code can be obtained with the following command:
\begin{lstlisting}[language=bash]
$ grep "TIME TO FORM FOCK" <logfile> \
	| awk '{print $9}'
\end{lstlisting}
It will return the time in seconds for the Fock matrix construction step.

\begin{table}[ht]
	\centering
  	\caption{Key configuration parameters of the install.info file for the supercomputers used in this work}
  	\label{tab:ad:clust}
  	\begin{tabular}{lll}
    \toprule
    Parameter 			& JLSE		& Cray XC40 \\
    \midrule
    GMS\_TARGET			&	linux64	&	cray-xc	\\
    GMS\_FORTRAN		&	ifort	&	ftn		\\
	GMS\_MATHLIB		&	mkl		&	mkl 	\\
    GMS\_DDI\_COMM		&	mpi		&	mpi		\\
    GMS\_MPI\_LIB		&	impi	&	impi	\\
    GMS\_LIBCCHEM		&	false	&	false	\\
    GMS\_PHI			&	true	&	true	\\
    GMS\_SHMTYPE		&	sysv	&	posix	\\
    GMS\_OPENMP			&	true	&	true	\\
  	\bottomrule
\end{tabular}
\end{table}

\subsection{Experiment customization}
During the experiment we changed the following runtime parameters:
\begin{itemize}
	\item The total number of MPI processes with batch script parameters
    \item The number of OpenMP threads per MPI process by adjusting \verb|OMP_NUM_THREADS| environmental variable
    \item Memory and clustering modes of \intelphireg\ nodes (BIOS parameters, restart is required)
    \item The environmental variables for Intel MPI and Intel OpenMP libraries (\verb|KMP_AFFINITY|, \verb|I_MPI_PIN_DOMAIN|)
\end{itemize}
\subsection{Notes}
Some of the GAMESS timer routines use CPU time instead of wall clock time without informing the user about it. Timing results from these routines will be incorrect for multithreaded code. Therefore, we used  \verb|omp_get_wtime()| function to measure the performance of OpenMP code.


\end{document}